\documentclass[twocolumn,showpacs,preprintnumbers,floatfix,
               amsmath,amssymb]{revtex4}
\usepackage{longtable}
\usepackage{graphics}
\usepackage{epsfig}

\begin{document}

\title{Unmasking the nuclear matter equation of state}

\author{J. Piekarewicz}

\affiliation{Department of Physics, Florida State University,
Tallahassee, Florida 32306}

\pacs{24.10.Jv, 21.10.Re, 21.60.Jz}

\date{\today}

\begin{abstract}
 Accurately calibrated (or ``best fit'') relativistic mean-field
 models are used to compute the distribution of isoscalar monopole
 strength in ${}^{90}$Zr and ${}^{208}$Pb, and the isovector dipole
 strength in ${}^{208}$Pb using a continuum random-phase-approximation
 approach. It is shown that the distribution of isoscalar monopole
 strength in ${}^{208}$Pb---but not in ${}^{90}$Zr---is sensitive to
 the density dependence of the symmetry energy. This sensitivity
 hinders the extraction of the compression modulus of symmetric
 nuclear matter from the isoscalar giant monopole resonance (ISGMR) in
 ${}^{208}$Pb. Thus, one relies on ${}^{90}$Zr, a nucleus with both a
 small neutron-proton asymmetry and a well developed ISGMR peak, to
 constrain the compression modulus of symmetric nuclear matter 
 to the range $K\!=\!(248\pm 6)~{\rm MeV}$. In turn, the sensitivity
 of the ISGMR in ${}^{208}$Pb to the density dependence of the
 symmetry energy is used to constrain its neutron skin to the range
 $R_{n}\!-\!R_{p}\!\alt\!0.22$~fm. The impact of this result on the
 enhanced cooling of neutron stars is briefly addressed.
\end{abstract}
\maketitle
%
Constraining the equation of state (EoS) of neutron-rich matter
remains a fundamental problem in nuclear physics and astrophysics.
The stability of neutron-rich nuclei~\cite{Tod03_PRC67}, the dynamics 
of heavy-ion collisions~\cite{Dan02_Science298,Li02_PRL88}, the
structure of neutron stars~\cite{Lat01_APJ550}, and the simulation of 
core-collapse supernova~\cite{Sum95_AA303,Bur03_PRL90},
all depend sensitively on the EoS. Unfortunately, our window to the
EoS is limited by terrestrial experiments that have, until now, probed
stable nucleonic matter at (or close to) nuclear-matter saturation
density. Fortunately, dramatic improvements are unfolding on several
fronts. First, the commissioning of new radioactive-beam facilities
all over the world will probe the EoS at large neutron-proton
asymmetries. By defining the limits of nuclear existence, these exotic
nuclei will constrain the EoS of neutron-rich matter at (and below)
normal nuclear densities. Second, space-based telescopes have started
to place important constrains on the high-density component of the
EoS~\cite{Pon02_APJ564,Wal02_APJ576}. New telescopes operating at a 
variety of wavelengths are turning neutron stars from theoretical 
curiosities into powerful diagnostic tools.

The nuclear matter equation of state is conveniently parametrized
in terms of the energy of symmetric nuclear matter (${\cal B/A}$) 
and the symmetry energy (${\cal S/A}$) in the following form:
\begin{eqnarray}
 && {\cal E/A}(k_{\rm F},b) - M
      = {\cal B/A}(k_{\rm F}) + b^{2} {\cal S/A}(k_{\rm F}) 
      +  {\cal O}(b^{4}) = \nonumber \\
     && \left(\epsilon_{0} \!+\! 
     \frac{1}{2}K\xi^{2} \!+\! \ldots\right) \!+\!
     b^{2}\left(J + L\xi \!+\! 
     \frac{1}{2}K_{\rm sym}\xi^{2}\!+\!\ldots\right).
\label{Eos}
\end {eqnarray}
Here the deviation from the equilibrium Fermi momentum is denoted 
by $\xi\!\equiv\!(k_{\rm F}-k_{\rm F}^{0})/k_{\rm F}^{0}$, the
neutron-proton asymmetry by $b\equiv(N-Z)/A$, and the various
coefficients ($K$, $J$, $L$, $K_{\rm sym}$) parametrize the density
dependence of the EoS around saturation density. 

Seven decades of nuclear physics have placed important constraints on
the nuclear matter equation of state. Indeed, the energy systematics
of medium to heavy nuclei, when combined with accurately calibrated
models, place the saturation point of symmetric nuclear matter at a
density of $\rho_{0}\!\simeq\!0.15~{\rm fm}^{-3}$
($k_{F}^{0}\!\simeq\!1.3~{\rm fm}^{-1}$) and a binding-energy per
nucleon of $\epsilon_{0}\!\simeq\!-16$~MeV. It should be noted that
one of the main virtues of the above Taylor-series expansion around
saturation density [Eq.~(\ref{Eos})] is that the linear term in $\xi$
for symmetric nuclear matter ({\it i.e.,} the pressure) automatically
vanishes.  Yet no such {\it special} saturation point exists in the
case of the symmetry energy. Indeed, the symmetry energy at saturation
density is not well known. Rather, it is the symmetry energy at the
lower density of $\widetilde{\rho}_{0}\!\simeq\!0.10~{\rm fm}^{-3}$
($\widetilde{k}_{F}^{0}\!\simeq\!1.15~{\rm fm}^{-1}$) that seems to be
accurately constrained (to within 1 MeV) by available ground-state
observables~\cite{Bro00_PRL85,Fur02_NPA706}. It should be emphasized
that present-day experiments can fix only one {\it isovector}
quantity. If one insists---and one should not---on constraining the
parameters of the symmetry energy at saturation density, then one
would find a strong correlation among its parameters ($J$, $L$,
$K_{\rm sym}$, $\ldots$)~\cite{Fur02_NPA706}. For example,
relativistic models consistently predict larger values for both the
symmetry-energy coefficient $J$ and the slope $L$ at saturation
density relative to nonrelativistic Skyrme models. This must be so if
all models are to reproduce the value of the symmetry energy at the
lower Fermi momentum of $\widetilde{k}_{F}^{0}$. Thus, in the present
contribution we adopt the following convention: the symmetry energy is
expanded around $\widetilde{k}_{F}^{0}\!\simeq\!1.15~{\rm fm}^{-1}$
and the value of the symmetry energy at $\widetilde{k}_{F}^{0}$ is
fixed at $\widetilde{J}\!=\!25.67$~MeV. That is,
\begin{equation}
  {\cal S/A}(k_{\rm F}) =
  \widetilde{J} + \widetilde{L}\tilde{\xi} + 
  \frac{1}{2}\widetilde{K}_{\rm sym}\tilde{\xi}^{2}  
  + \ldots
  \label{SymmE}
\end{equation}
where 
$\tilde{\xi}\!=\!(k_{\rm F}-\widetilde{k}_{\rm F}^{0})/
\widetilde{k}_{\rm F}^{0}$. Note that henceforth ``tilde''
quantities refer to parameters of the symmetry energy at
$\widetilde{k}_{F}^{0}\!\simeq\!1.15~{\rm fm}^{-1}$.

Having established that existing ground-state observables accurately
determine the binding energy per nucleon $\epsilon_{0}$ at $k_{F}^{0}$
and the symmetry-energy coefficient $\widetilde{J}$ at
$\widetilde{k}_{F}^{0}$, how can one constrain any further the density
dependence of the equation of state?  In the case of symmetric nuclear
matter, the dynamics of small density fluctuations around the
saturation point is controlled by the compression modulus $K$. The
isoscalar giant monopole resonance (ISGMR) in heavy nuclei has long
been regarded as the optimal observable from which to determine the
compression modulus~\cite{Bla80_Pr64}. This is especially true now
that the {\it breathing mode} has been measured on a variety of nuclei
with unprecedented accuracy~\cite{You99_PRL82}. In contrast, the
density dependence of the symmetry energy is poorly
constrained. Indeed, one may fit a variety of ground-state observables
(such as charge densities, binding energies, and single-particle
spectra) using accurately-calibrated models that, nevertheless,
predict a wide range of values for the neutron skin of
${}^{208}$Pb~\cite{Hor01_PRL86}.  As the neutron skin of a heavy
nucleus is strongly correlated to the slope of the symmetry
energy~\cite{Bro00_PRL85,Fur02_NPA706}, measuring the skin thickness
of a single heavy nucleus will constrain the density dependence of the
symmetry energy. The Parity Radius Experiment (PREX) at the Jefferson
Laboratory aims to measure the neutron radius of $^{208}$Pb accurately
(to within $0.05$~fm) and model independently via parity-violating
electron scattering~\cite{Hor01_PRC63,Mic02_JLAB003}. This experiment
should provide a unique observational constraint on the density
dependence of the symmetry energy.

While the above arguments suggest a clear path toward constraining the
density dependence of the EoS, theoretical uncertainties have clouded
these issues. First and foremost is the apparent discrepancy between
nonrelativistic and relativistic predictions for the value of the
compressional modulus of symmetric nuclear matter required to
reproduce the ISGMR in $^{208}$Pb. While nonrelativistic models
predict
$K\!\simeq\!220\!-\!235$~MeV~\cite{Col92_PLB276,Bla95_NPA591,Ham97_PRC56},
relativistic models argue for a significantly larger value
$K\!\simeq\!250\!-\!270$~MeV~\cite{Lal97_PRC55,Vre00_PLB487,Vre03_PRC68}.
Further, relativistic models systematically predict larger values for
the neutron skin of $^{208}$Pb relative to nonrelativistic Skyrme
models. One of the goals of this contribution is to show that these
two points are related. Indeed, the aim of this contribution is
twofold.  First, to vindicate---through the exclusive use of
accurately-calibrated models---our previous assertion that the
distribution of ISGMR in heavy nuclei, and therefore the inferred
value of $K$, is sensitive to the density dependence of the symmetry
energy~\cite{Pie02_PRC66}.  Second, to rely on existing data on the
isoscalar giant-monopole resonance in ${}^{90}$Zr and
${}^{208}$Pb~\cite{You99_PRL82}, and on the isovector giant-dipole
resonance in ${}^{208}$Pb~\cite{Rit93_PRL70}, to set
limits---simultaneously---on the compression modulus of symmetric
nuclear matter and on the neutron skin of ${}^{208}$Pb. Note that
since first proposed~\cite{Pie02_PRC66}, other groups have addressed
the possible impact of the density dependence of the symmetry energy 
on the ISGMR in
${}^{208}$Pb~\cite{Vre03_PRC68,Agr03_PRC68,Col03_NuclTh}.

The starting point for the calculations is an interacting Lagrangian
density of the following form~\cite{Hor01_PRL86,Hor01_PRC64}:
\begin{eqnarray}
 &&{\mathcal L}_{\rm int} \!=\! \bar{\psi}\left[
   g_{\rm s}\phi\!-\!\left(g_{\rm v}V_{\mu}\!+\!\frac{g_{\rho}}{2}
   {\mbox{\boldmath$\tau$}} \cdot {\bf b_{\mu}}\!+\!
   \frac{e}{2}(1+\tau_{3})A_{\mu}\right)\!\gamma^\mu\right]\psi
   \nonumber \\ && \qquad 
       - \frac{\kappa}{3!}\Phi^{3}
       -  \frac{\lambda}{4!}\Phi^{4}
       +  \Lambda_{\rm v}\left(W_{\mu}W^\mu\right)
          \left({\bf B}_{\mu}\cdot{\bf B}^\mu\right)\;,
 \label{Lint}
\end{eqnarray}
where $\Phi\!=\!g_{\rm s}\phi$, $W_{\mu}\!=\!g_{\rm v}V_{\mu}$, and
${\bf B}_{\mu}\!=\!g_{\rho}{\bf b}_{\mu}$. The Lagrangian density
includes Yukawa couplings of the nucleon field to a scalar ($\phi$)
and to three vector ($V^{\mu}$, ${\bf b}^{\mu}$, and $A^{\mu}$) fields.
In addition to the Yukawa couplings, the Lagrangian is supplemented
by three nonlinear meson interactions. The inclusion of scalar-meson
interactions (via $\kappa$ and $\lambda$) is used to soften the 
equation of state of symmetric nuclear matter, while the mixed
isoscalar-isovector coupling ($\Lambda_{\rm v}$) modifies the density
dependence of the symmetry energy---without affecting well known
ground-state properties. Note that this last term was absent from 
Ref.~\cite{Pie02_PRC66}, so the softening of the symmetry energy 
had to be done artificially. This drawback has now been corrected.

The relativistic mean-field models employed here are motivated by the
enormous success of the NL3
parametrization~\cite{Lal97_PRC55}. Indeed, the NL3000 set used here
(having $\Lambda_{\rm v}\!=\!0$) is practically identical to the
original NL3 model. The other sets are obtained by adding an
isoscalar-isovector coupling $\Lambda_{\rm v}\!\ne\!0$, while at the
same time re-adjusting the strength of the $NN\rho$ coupling constant
($g_\rho$) to maintain the symmetry-energy coefficient fixed at
$\widetilde{J}\!=\!25.67$~MeV (see discussion above). The aim of this
added coupling is to change the neutron density of heavy nuclei, while
leaving intact ground-state observables that are well constrained
experimentally.  One should stress that the addition of $\Lambda_{\rm
v}$ has no impact on the properties of symmetric nuclear matter, so
the saturation properties remain unchanged. In summary, all the models
used in this contribution share the following properties with the
original NL3 model of Ref.~\cite{Lal97_PRC55}: for symmetric nuclear
matter, a Fermi momentum at saturation of $k_{F}^{0}\!=\!1.30~{\rm
fm}^{-1}$ with a binding-energy per nucleon of
$\epsilon_{0}\!=\!-16.24$~MeV, and a compression modulus of
$K\!=\!271$~MeV. For the symmetry energy, a symmetry-energy
coefficient of $\widetilde{J}\!=\!25.67$~MeV at a Fermi momentum of
$\widetilde{k}_{F}^{0}\!=\!1.15~{\rm fm}^{-1}$.

\vspace{0.25in}
\begin{figure}[ht]
\begin{center}
\includegraphics[width=3.0in,angle=0,clip=false]{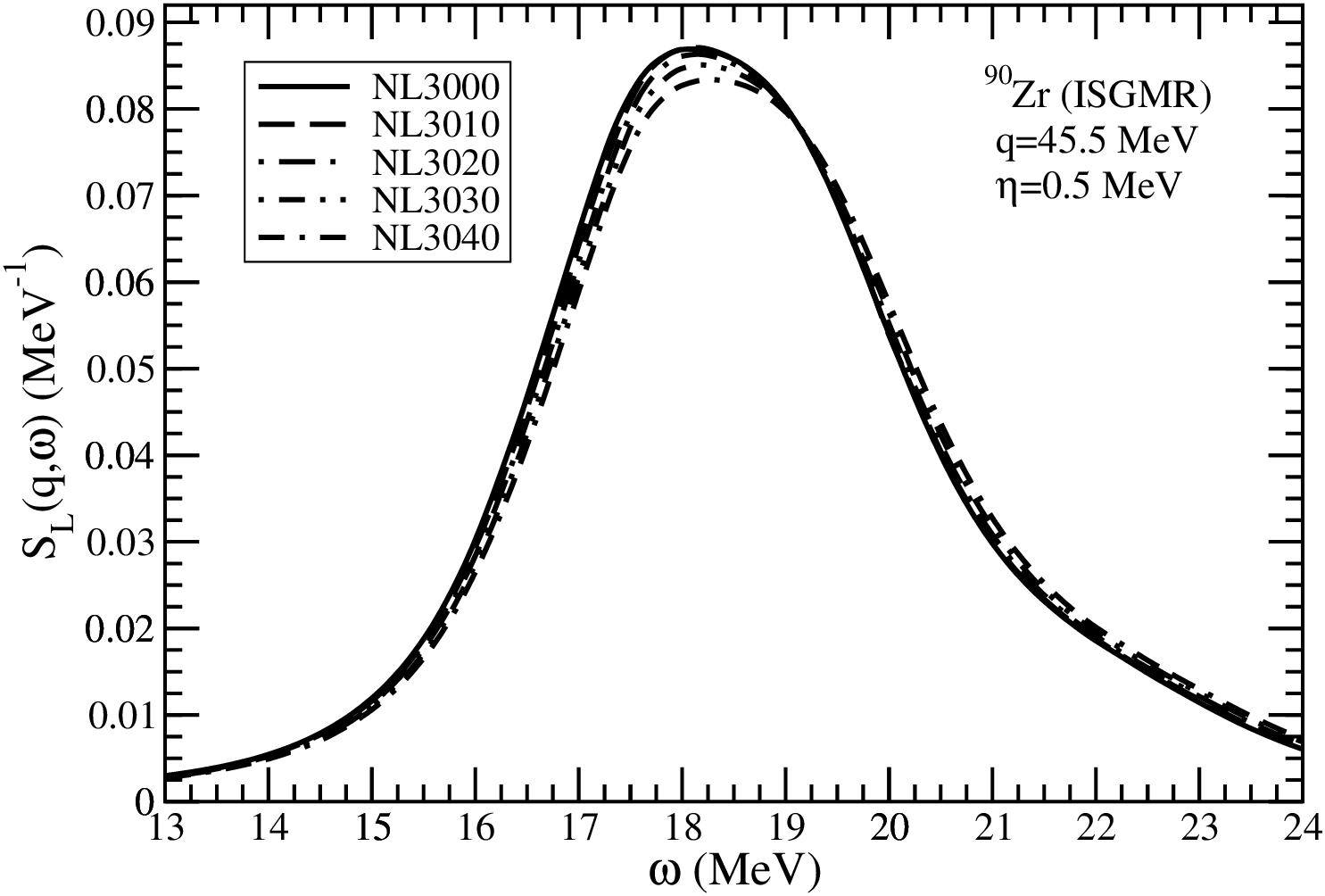}
\caption{Distribution of isoscalar monopole strength in
         ${}^{90}$Zr for the small momentum transfer of
         $q\!=\!45.5$~MeV. The response includes a small 
         artificial width of $0.5$~MeV.}  
\label{Fig1}
\end{center}
\end{figure}

\begin{figure}[ht]
\begin{center}
\includegraphics[width=3.0in,angle=0,clip=false]{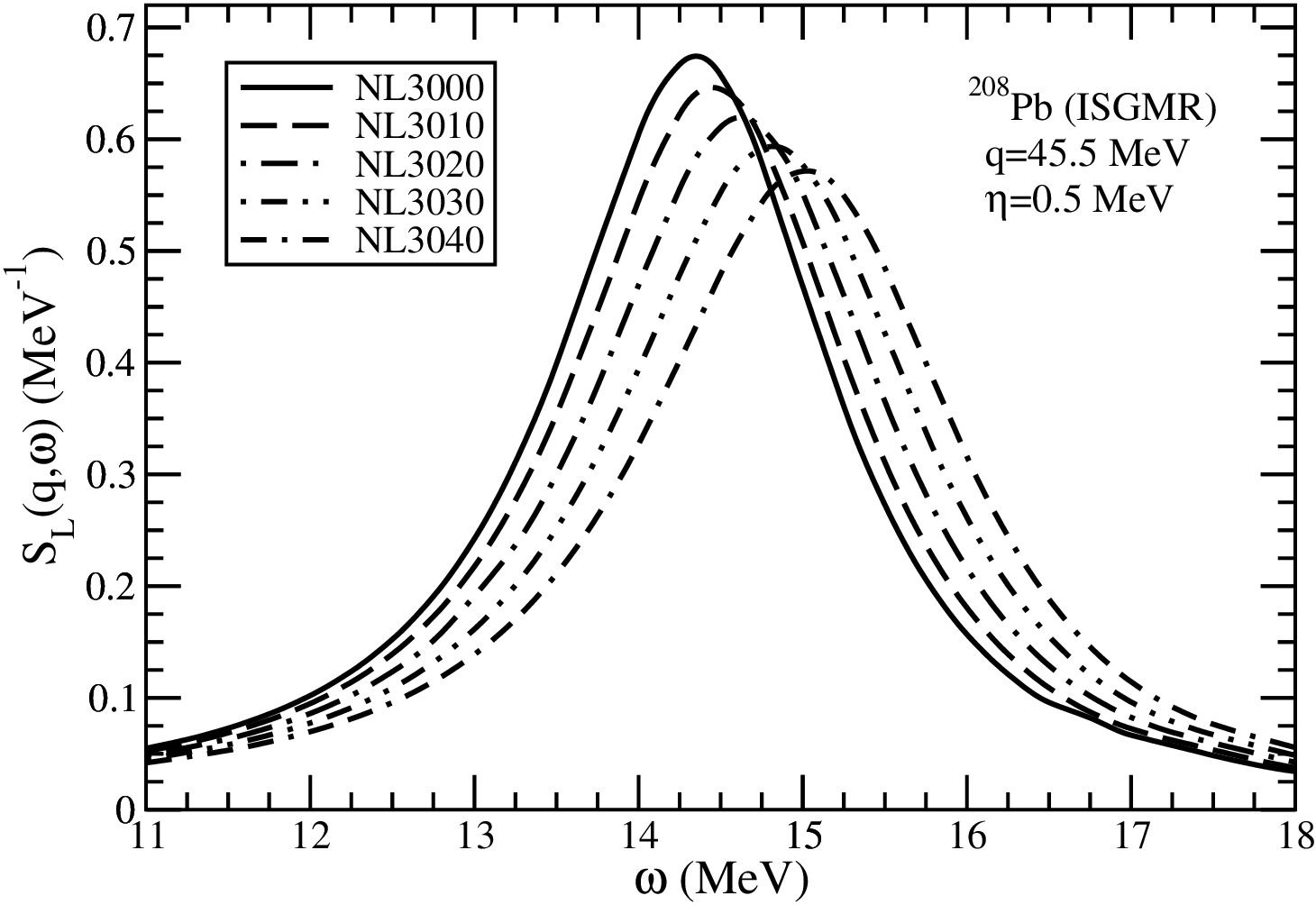}
\caption{Distribution of isoscalar monopole strength in
         ${}^{208}$Pb for the small momentum transfer of
         $q\!=\!45.5$~MeV. The response includes a small 
         artificial width of $0.5$~MeV.}  
\label{Fig2}
\end{center}
\end{figure}

While the success of the NL3 interaction in reproducing ground-state
properties (such as binding energies, charge radii, energy
separations, {\it etc.}) for a variety of nuclei all throughout the
periodic table is well documented, we display in Table~\ref{table1}
ground-state properties for only the two nuclei of relevance to this
contribution, namely, ${}^{90}$Zr and ${}^{208}$Pb. However, even
these accurately calibrated models predict a wide range of values for
the neutron skin of ${}^{208}$Pb, confirming that the neutron-skin of
a heavy nucleus is not tightly constrained by known nuclear
observables. The fifth column in the table displays the compression
modulus of {\it asymmetric} nuclear matter with a neutron-proton
asymmetry corresponding to ${}^{90}$Zr ($b\!=\!0.111$) and
${}^{208}$Pb ($b\!=\!0.212$). It is this quantity---not the
compression modulus of symmetric nuclear matter---that is constrained
by the breathing mode of nuclei. This simple fact makes the connection
between the measured ISGMR and the compression modulus of {\it symmetric}
nuclear matter sensitive to the density dependence of the symmetry
energy. Recall that the compression modulus of symmetric nuclear
matter was fixed in all models at $K\!=\!271$~MeV, yet for
($b\!=\!0.212$) asymmetric nuclear matter the compression
modulus ranges from $243$~MeV (for the stiffest symmetry energy) all
the way up to $260$~MeV (for the softest symmetry energy). Finally,
the last column in the table shows peak and centroid energies for the
ISGMR in ${}^{90}$Zr and ${}^{208}$Pb computed in a relativistic
random-phase-approximation (RPA); Figs.~\ref{Fig1} and~\ref{Fig2}
display the corresponding distribution of strength. As expected, there
is a strong correlation between the centroid energies and the
compression modulus of {\it asymmetric} nuclear matter. One should
note in passing that the continuum RPA formalism employed here, but
reported elsewhere~\cite{Pie00_PRC62,Pie01_PRC64}, respects important
symmetries of nature, such as translational invariance (in the form of
Thouless' theorem~\cite{Tho61_NP22,Daw90_PRC42}) and the conservation 
of the vector current.

\vspace{0.25in}

\begin{table*}
\caption{Binding energy per nucleon, root-mean-square charge radius,
         neutron-minus-proton root-mean-square radius, compression
	 modulus for asymmetric ($b\!=\!0.111$ and) nuclear matter, 
	 and peak and centroid GMR energies for ${}^{90}$Zr in the 
	 various models discussed in the text. The binding energy 
	 includes a center-of-mass correction of $-0.08$~MeV/nucleon,
	 while the centroid energy ($m_{1}/m_{0}$) was computed by
         generating the distribution of strength in the range 
	 $10\!\le\!\omega\!\le\!26$~MeV. The second set of numbers 
	 in the table are for ${}^{208}$Pb ($b\!=\!0.212$) with a
	 center-of-mass correction of $-0.02$~MeV/nucleon and a
	 centroid energy extracted from a distribution of strength
	 generated in the range $8\!\le\!\omega\!\le\!23$~MeV.}
 \label{table1}
 \begin{ruledtabular}
 \begin{tabular}{ccccccc}
  Model & $B/A$~(MeV) & $r_{\rm ch}$~(fm) & $R_{n}\!-\!R_{p}$~(fm)
        & $K_{b}$~(MeV) & $E_{\rm GMR}~[m_{1}/m_{0}]$~(MeV) \\
  \hline
  NL3000 & 8.69 & 4.26 & 0.11 & 263.13 & 18.10 [18.62] \\ 
  NL3010 & 8.69 & 4.26 & 0.10 & 263.76 & 18.14 [18.67] \\ 
  NL3020 & 8.70 & 4.26 & 0.09 & 265.23 & 18.15 [18.69] \\ 
  NL3030 & 8.70 & 4.27 & 0.08 & 266.84 & 18.20 [18.75] \\ 
  NL3040 & 8.70 & 4.27 & 0.07 & 268.32 & 18.25 [18.77] \\ 
  \hline
  Experiment & 8.71$\pm$0.01 & 4.26$\pm$0.01 & 
             unknown & unknown & [17.89$\pm$0.20] \\ 
  \hline
  \hline
  NL3000 & 7.87 & 5.51 & 0.28 & 242.93 & 14.35 [14.32] \\ 
  NL3010 & 7.89 & 5.51 & 0.25 & 244.22 & 14.45 [14.43] \\ 
  NL3020 & 7.91 & 5.51 & 0.22 & 248.88 & 14.62 [14.57] \\ 
  NL3030 & 7.91 & 5.52 & 0.20 & 254.46 & 14.82 [14.74] \\ 
  NL3040 & 7.92 & 5.53 & 0.17 & 259.87 & 15.03 [14.91] \\ 
  \hline
  Experiment & 7.87$\pm$0.01 & 5.50$\pm$0.01 & 
             unknown & unknown & [14.24$\pm$0.11] \\ 
 \end{tabular}
\end{ruledtabular}
\end{table*}

\begin{figure}[ht]
\begin{center}
\includegraphics[width=3.0in,angle=0,clip=false]{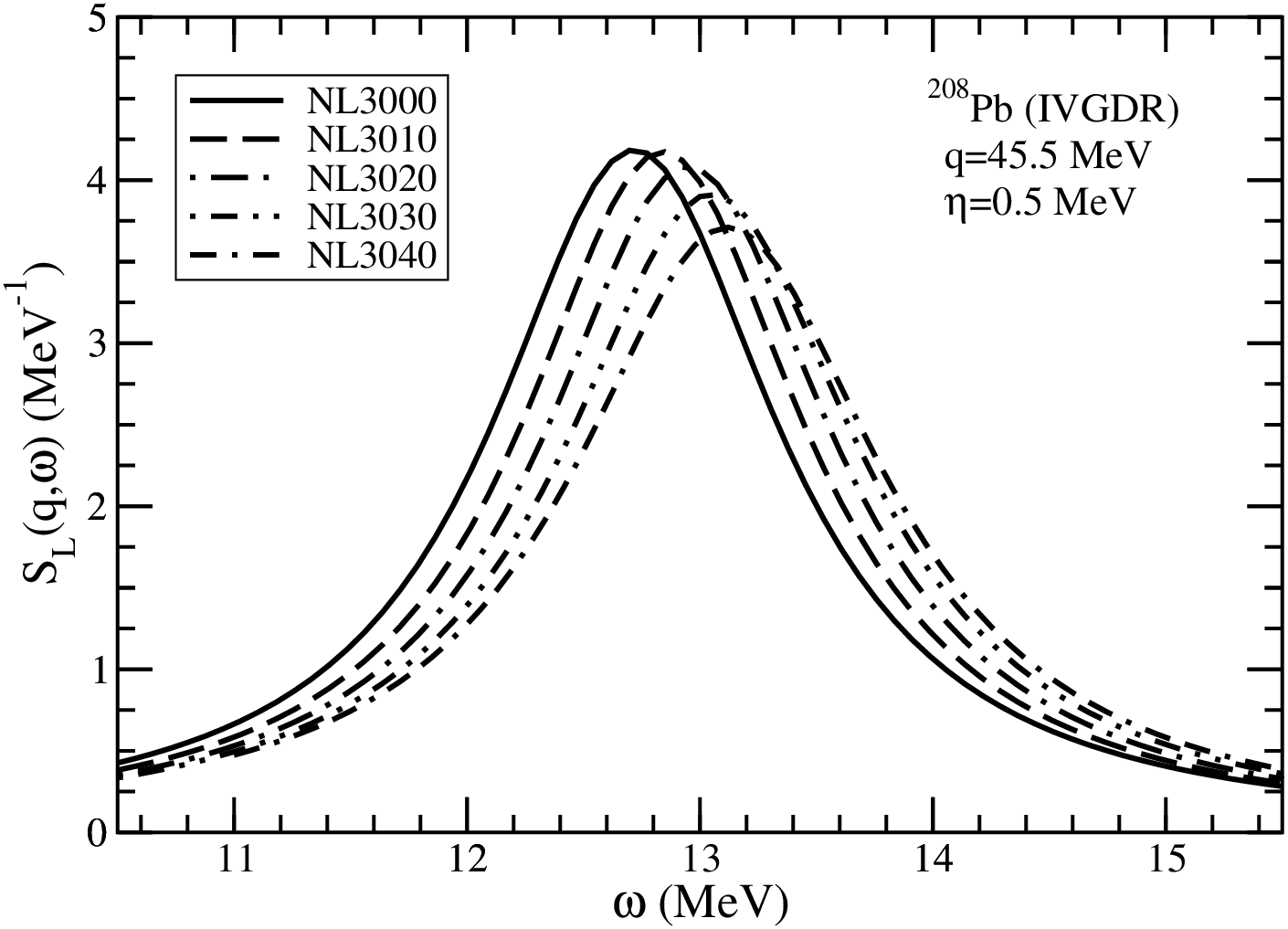}
\caption{Distribution of isovector dipole strength in
         ${}^{208}$Pb for the small momentum transfer of
         $q\!=\!45.5$~MeV. The response includes a small 
         artificial width of $0.5$~MeV.}  
\label{Fig3}
\end{center}
\end{figure}

The great advantage of a nucleus such as ${}^{90}$Zr is that it has
both a well developed isoscalar-monopole peak and a small
neutron-proton asymmetry ($b\!=\!0.111$). The latter manifests itself
into the near collapse of all curves in Fig.~\ref{Fig1} into a single
one, so that the former may directly constrain the compression modulus
of symmetric nuclear matter. In contrast to ${}^{90}$Zr, {\it the
distribution of ISGMR strength in ${}^{208}$Pb is sensitive to the
density dependence of the symmetry energy.} While this sensitivity
should be sufficient to constrain the density dependence of the
symmetry energy, one could do even better. Indeed, one may constrain
the density dependence of the symmetry energy by demanding that both
the ISGMR and the isovector giant-dipole resonance (IVGDR) in
${}^{208}$Pb be simultaneously reproduced. The distribution of
isovector dipole strength in ${}^{208}$Pb is displayed in
Fig.~\ref{Fig3}. We note that the isovector-dipole response gets
hardened as the symmetry energy is softened. As all models share the
same value of the symmetry-energy coefficient at
$\widetilde{k}_{F}^{0}\!=\!1.15~{\rm fm}^{-1}$, the hardening of the
response follows as a result of the symmetry energy being higher at
the (low) densities relevant to the isovector-dipole
mode~\cite{Vre03_PRC68}.

\vspace{0.2in}

\begin{figure}[ht]
\begin{center}
\includegraphics[width=3.0in,angle=0,clip=false]{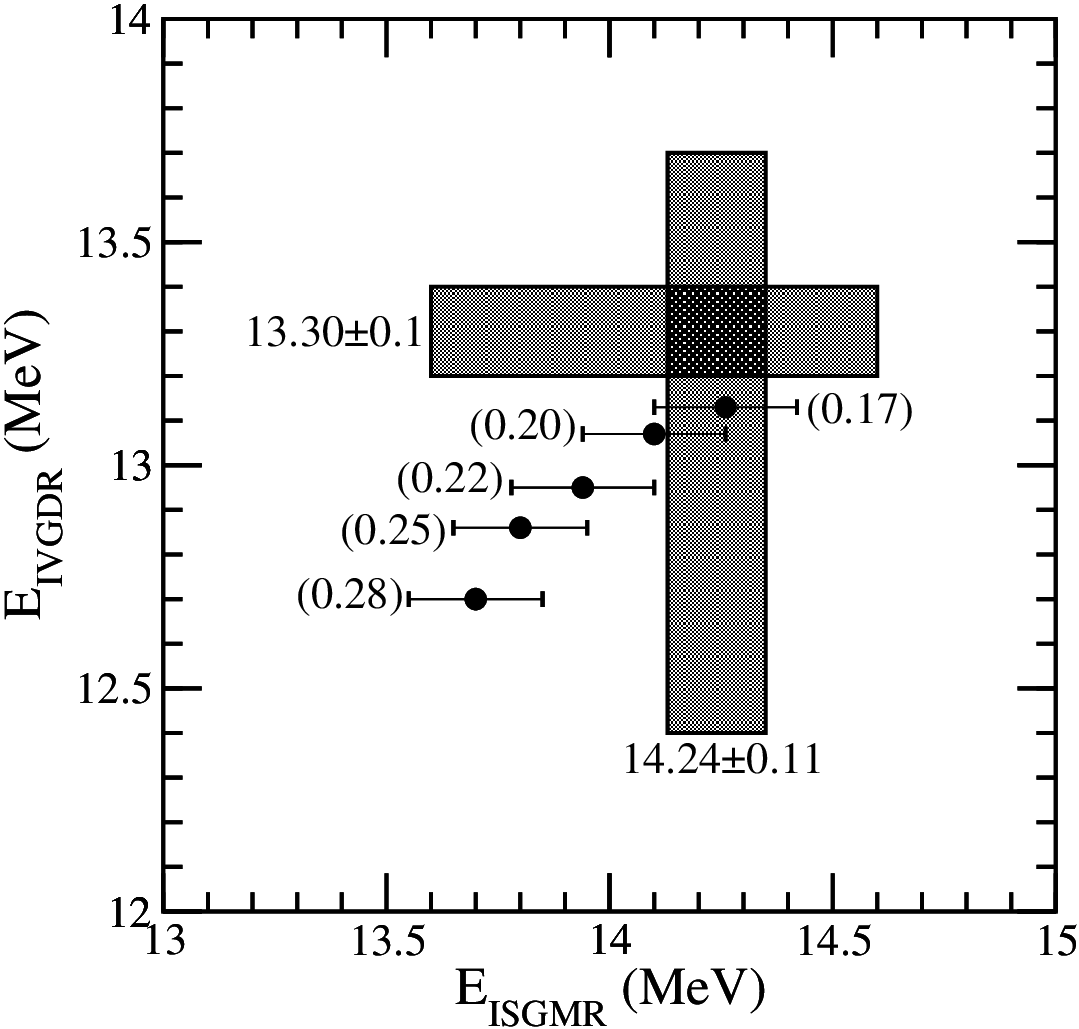}
\caption{Comparison between theoretical and experimental 
         ISGMR centroid and IVGDR peak energies for 
         ${}^{208}$Pb. Quantities in parenthesis represent
	 the predictions for the neutron-skin of ${}^{208}$Pb
	 in the various models discussed in the text.}
\label{Fig4}
\end{center}
\end{figure}

To constrain simultaneously the compression modulus of symmetric
nuclear matter and the neutron radius of ${}^{208}$Pb, one starts by
noticing that the theoretical centroid energy of the ISGMR in
${}^{90}$Zr overestimates the experimental value by about
$1$~MeV. Although a proper adjustment of $K$ should be done through a
re-calibration of parameters, a simple, yet accurate estimate may be
obtained via the following scaling relation: $E_{\rm
ISGMR}\propto\sqrt{K}$~\cite{Bla95_NPA591}. Using this relation and
accounting for experimental uncertainties, an adjustment of about
$20$~MeV in $K$ is required to reproduce the ISGMR in
${}^{90}$Zr. That is, $K\!=\!271~{\rm MeV}\rightarrow K\!=\!(248\pm
6)~{\rm MeV}$. This adjustment in $K$ induces a corresponding
correction in the calculated values of the ISGMR in ${}^{208}$Pb. The
centroid energies after correction, together with the peak energies of
the IVGDR in ${}^{208}$Pb, are displayed in Fig.~\ref{Fig4}, alongside
the experimental values~\cite{You99_PRL82,Rit93_PRL70}. The numbers in
parenthesis indicate the predicted values for the neutron skin in
${}^{208}$Pb. The figure suggests that models with neutron skins in
${}^{208}$Pb larger than $R_{n}\!-\!R_{p}\simeq 0.22$~fm may have an
unrealistically stiff symmetry energy.

\vspace{0.25in}

In summary, relativistic mean-field models have been used to compute 
the distribution of isoscalar-monopole strength in ${}^{90}$Zr and 
${}^{208}$Pb, and of isovector-dipole strength in ${}^{208}$Pb using 
a continuum RPA approach. It was demonstrated---using exclusively 
accurately calibrated models---that the distribution of isoscalar 
monopole strength in ${}^{208}$Pb is sensitive to the density
dependence of the symmetry energy. Further, existing experimental 
data were used to set limits on both the compression modulus of 
symmetric nuclear matter and on the neutron skin of ${}^{208}$Pb. 
It appears that medium-mass nuclei, having a well-developed ISGMR 
peak and a small neutron-proton asymmetry (such as ${}^{90}$Zr but 
not ${}^{208}$Pb) allow for the best determination of the compression 
modulus of symmetric nuclear matter. In turn, the sensitivity of the 
ISGMR and the IVGDR in ${}^{208}$Pb to the density dependence of the 
symmetry energy may be used to impose constraints on the neutron skin 
of ${}^{208}$Pb. From the present analysis, a compression modulus of 
$K\!=\!(248\pm 6)~{\rm MeV}$ and a neutron skin in ${}^{208}$Pb of 
$R_{n}\!-\!R_{p}\alt 0.22$~fm were obtained. These values appear
closer to those predicted in nonrelativistic studies.

We conclude with a comment on the impact of these results on the
cooling of neutron stars. In earlier publications we have demonstrated
how improved values for neutron radii could have a widespread impact
on the structure and dynamics of 
neutron stars~\cite{Hor01_PRL86,Hor01_PRC64,Hor02_PRC66,Car03_APJ593}.
In particular, we suggested that the enhanced cooling of the neutron 
star in 3C58~\cite{Sla02_APJ571} may be due to the conventional URCA 
process---provided the symmetry energy is stiff enough to generate a 
neutron skin in ${}^{208}$Pb larger than $R_{n}\!-\!R_{p}\agt
0.24$~fm. In view of our present findings, this now seems unlikely. 
Thus, the possibility that 3C58 harbors an exotic star, such as a 
quark star, looms large.

\begin{acknowledgments}
\vspace{-0.15in}
The author is grateful to the organizers of the International 
Workshop on {\it Nuclear Response under Extreme Conditions}
in Trento and to the ECT* for their support and hospitality. 
This work was supported in part by the U.S. Department of 
Energy under Contract No.DE-FG05-92ER40750.
\end{acknowledgments}
\vfill\eject

\bibliography{GMR}
\end{document}